\documentclass[a4paper,superscriptaddress,showkeys,prX,showpacs,twocolumn]{revtex4-1}
\usepackage{amssymb, amsmath, amsfonts,tensor,bm}
\usepackage{graphics}
\usepackage{pst-all}
\usepackage{hyperref}
\usepackage[mathlines,running,modulo]{lineno}
\usepackage{relsize}
\usepackage[utf8]{inputenc}
\usepackage{IEEEtrantools}
\usepackage{epsfig}
\usepackage[toc]{appendix}

\begin{document}
\title{Effectiveness of the Self-Consistent Harmonic Approximation in ferromagnets with dipolar interactions}
\author{A. R. Moura}
\email{antoniormoura@ufv.br}
\affiliation{Departamento de F\'{i}sica, Universidade Federal de Vi\c{c}osa, 36570-900, Vi\c{c}osa, Minas Gerais, Brazil}
\date{\today}
\begin{abstract}
Among the various methods for treating magnetic models, the Self-Consistent Harmonic Approximation (SCHA) has successfully 
described ferro and antiferromagnetism in many different scenarios. In particular, the SCHA is a valuable
and easy formalism for determining transition temperatures as, for example, the Berezinskii-Kosterlitz-Thouless. The heart of 
the method includes thermal fluctuations through of a renormalization parameter depending on temperature. Nevertheless, most of 
the work has been done considering only short-range interactions, which results in an incomplete description of actual magnetic samples. 
Here, we generalize the SCHA to include the dipolar interaction in the thermodynamic analysis. The method is applied 
to analyze the well-known Europium Chalcogenides EuO and EuS. The SCHA results are in good agreement with the experimental measurements.
\end{abstract}
  

\maketitle

\section{Introduction and motivation}

The description of magnetism in condensed matter physics involves a diversified set of 
theoretical tools. For a long time, the bosonic representations, for example, have been widely used to investigate
all kinds of magnetic properties, spin excitations, and phase transitions in ferro (FM) and antiferromagnetic (AFM) models. The main
concept is the replacement of the spin operators by annihilation/creation bosonic ones. Since there are many
bosonic representations, one should choose the more appropriate formalism according to the model's dimensionality, temperature, spin
interactions, and/or symmetries. At low temperatures (below the ordering temperature), it is usual to adopt the 
Holstein-Primakoff representation \cite{pr58.1098} since the spontaneous symmetry breaking justifies the series expansion of 
the spin operators in lower orders of the magnon occupation $n=a^\dagger a\ll 1$ \cite{auerbach}. In the lowest order, we have
the traditional linear spin-wave theory, which is a reasonable picture of magnons weakly coupled. On the other hand, phases with 
intact symmetry are better described by using the Schwinger bosonic representation \cite{prb38.316, prb40.5028, jap67.5734, auerbach}, 
although three-dimensional models require special attention close to the transition temperature \cite{prb66.014407}.  
In general, the mean-field approach of the Schwinger formalism is sufficient for most of the scenarios; however, in frustrated
models, the inclusion of Gaussian fluctuations should be considered \cite{prl78.2216,prb96.174423,prb98.184403,prb100.104431}, 
providing some extra complexity to the model. Moreover, it is also possible to represent the spin field by the non-linear sigma 
model O(3)\cite{rajaraman,nagaosa,auerbach} and then quantize the field fluctuations by standard techniques of quantum field 
theory (furthermore, note that in the AFM case, one should be careful with the topological phase). In addition, the Self-Consistent
Gaussian Approximation (SCGA) \cite{prb53.11593} presents a purpose similar to the Self-Consistent Harmonic Approximation (SCHA). 
In the SCGA, the thermodynamics of a classical spin model is evaluated through self-consistent equations
depending on the magnetization and their quadratic fluctuations. In this case, the Gaussian corrections are introduced by 
considering spin cumulants \cite{pr124.1757,pr130.155} in the statistical averages. The SCGA formalism provides good results; 
however, the number of self-consistent parameters is larger than the SCHA, and the quantization is more challenging to implement.

The Self-Consistent Harmonic Approximation is another practical approach for solving spin models \cite{jp35.27}. 
Classically, the spin fields can be written using the phase angle $\varphi$ around the z-axis and the spin component $S^z$.
It is clear that $\varphi$ and $S^z$ composite a pair of canonically conjugate fields that obey the Poisson bracket 
$\{\varphi_i,S_j^z\}=\delta_{ij}$. In the quantum point of view, the development
is similar with the fields being replaced by operators that satisfy the commutation relation $[\varphi_i,S_j^z]=i\hbar \delta_{ij}$. 
Over the years, Pires {\it et al.} have applied the SCHA method to evaluate the 
critical temperature \cite{prb49.9663,pla202.309,prb51.16413,prb54.3019,ssc104.771,prb59.6229}, 
the topological BKT transition \cite{pla166.330,prb48.12698,prb49.9663,prb50.9592,ssc100.791,prb53.235,prb54.3019,prb54.6081,ssc112.705,epjb2.169,pssb.242.2138,prb78.212408,jmmm452.315}, 
and the large-D quantum phase transition \cite{pasma373.387,jpcm20.015208,pasma388.21,pasma388.3779,jmmm357.45} 
in a wide variety of magnetic models. Similar to the linear spin-wave theory, in the SCHA, we expand the Hamiltonian up to
second order in $\varphi$ and $S^z$. However, higher-order contributions are taken into account through a 
renormalization parameter $\rho$ that depends on temperature. The renormalization temperature is determined by a 
self-consistent equation that normally presents a fast convergence. Therefore, the SCHA method provides an easy and 
efficient alternative for investigating spin models, mainly in determining thermal and quantum phase transitions without 
the disadvantages of the usual bosonic formalisms. In addition, Moura and Lopes have demonstrated that, since $\varphi$ and $S^z$ 
are canonically conjugated, SCHA is the natural choice to describe coherent states in magnetic models \cite{jmmm472.1}, 
making SCHA very useful for describing the magnetization precession in spintronics.

Despite the success of the SCHA to describe spin models with short-range interactions, there are only a couple of works
considering Hamiltonians with long-range interactions. Pires investigated the phase transition of the Heisenberg
model with ferromagnetic long-range interaction decaying as $r^{-p}$ \cite{pla202.309}, and Moura generalized the results to 
the anisotropic Heisenberg model \cite{jmmm369.62}. In both cases, the long-range interaction was considered 
isotropic, limiting the possible applications. This paper, will present a complete development
of the SCHA that includes the dipolar interaction. As it is well-known, the dipolar interaction is an anisotropic
long-range interaction that decays as $r^{-3}$. In general, the dipolar interaction is weak, compared to the exchange
coupling, and insufficient to sustain an ordered phase by itself. However, the dipolar field has a fundamental role 
in the description of ferromagnetic insulators, as the ferromagnetic Yttrium-Iron-Garnet, which is commonly used in 
spintronics \cite{rezende}. Furthermore, to verify the obtained results, the SCHA is used to determine the 
thermodynamics of the ferromagnetic Europium Chalcogenides EuS and EuO \cite{wachter}. In both cases, the results agree
with the literature.

\section{The SCHA method}

We will consider a ferromagnetic insulator in a cubic crystalline lattice with an exchange interaction between nearest
neighbors. In the SI unit system, the Heisenberg Hamiltonian endowed with dipolar interaction is given by
\begin{IEEEeqnarray}{rCl}
\label{eq.hamiltonian}
H&=&-J\sum_{\langle ij\rangle}{\bf S}_i\cdot{\bf S}_j+\frac{\mu_0}{4\pi}\frac{(g\mu_B)^2}{2}\sum_{ij}\left[\frac{{\bf S}_i\cdot{\bf S}_j}{r_{ij}^3}-\right.\nonumber\\
&&\left.-3 \frac{({\bf S}_i\cdot{\bf r}_{ij})({\bf S}_j\cdot{\bf r}_{ij})}{r_{ij}^5}\right],
\end{IEEEeqnarray}
where the first sum is done over nearest neighbors and the second one is evaluated over each spin pair, separated by 
${\bf r}_{ij}={\bf r}_j-{\bf r}_i$, on the lattice; $J>0$ is the ferromagnetic exchange coupling, 
$\mu_0=4\pi\times 10^{-7}$ H/m is the vacuum permeability, $g$ is the Land\'{e} g-factor, and $\mu_B=9.274\times 10^{-24}$ J/T is 
the Bohr magneton. If necessary, other interactions, anisotropies, and coupling with magnetic fields can be easily included. 
For example, if one considers a finite model, it is important to include the surface contribution through the 
demagnetizating field, which vanishes for an ellipsoidal sample with magnetization oriented along a symmetry axis. 
Slight differences between theoretical and experimental results should be observed if minor effects are
disregarded. However, since our main objective is to investigate the dipolar interaction under the SCHA perspective, we consider 
only bulk interactions in this work. In above equation, ${\bf S}_i$ is adopted as a classical dimensionless spin field on the site $i$, whose transverse components are written as
\begin{IEEEeqnarray}{rCl}
\IEEEyesnumber
\IEEEyessubnumber*
S_i^x&=&\sqrt{S^2-(S_i^z)^2}\cos\varphi_i\\
S_i^y&=&\sqrt{S^2-(S_i^z)^2}\sin\varphi_i
\end{IEEEeqnarray}
In the quantum development, $\varphi_i$ and $S_i^z$ are promoted to operators that obey the usual commutation relation 
$[\varphi_i,S_j^z]=i\delta_{ij}$ \cite{jp35.27}. Notwithstanding some different procedures in the middle of the process, 
the final result obtained from the quantum version of the Hamiltonian (\ref{eq.hamiltonian}) is the same one obtained from
the {\it a posteriori} quantized Hamiltonian. Therefore, for the sake of simplicity, we will consider classical spin fields for now
and leave the quantization to be performed later. Let us define the magnetization along the $\langle 100\rangle$ direction and consider the fields $\varphi$ and $S^z$ small
enough to perform a series expansion around zero. After expanding up to second order in the $\varphi$ and $S^z$ fields, the 
exchange Hamiltonian reads in the momentum space
\begin{equation}
\label{eq.hamiltonianexc}
H_\textrm{exc}^{(2)}=\sum_q zJ(1-\gamma_q)(\rho S^2\varphi_q\varphi_{-q}+S_q^z S_{-q}^z ),
\end{equation}
where the $q$-sum extends over the first Brillouin zone. In above equation, the structure factor is defined by 
\begin{equation}
\gamma_q=\frac{1}{z}\sum_{{\bm \eta}}e^{i{\bf q}\cdot{\bm \eta}}, 
\end{equation}
where ${\bm \eta}$ stands for the nearest neighbor positions, and $z=6$ is the coordination number for the 
three-dimensional cubic lattice. In addition, $\rho$ is a renormalization parameter included to take into account 
higher-order phase fluctuations disregarded in the cosine expansion, which is determined by the self-consistent equation
\begin{equation}
\label{eq.rho}
\rho=\left[1-\frac{\langle (S^z)^2\rangle_0}{S^2}\right]\exp\left[-\frac{1}{2}\langle\Delta\varphi^2\rangle_0\right],
\end{equation}
where $\Delta\varphi$ is the phase difference between nearest neighbors, and the averages are evaluated by using the 
harmonic Hamiltonian $H_0=H_\textrm{exc}^{(2)}+H_\textrm{dip}^{(2)}$. More details about the renormalization parameter are 
given in Appendix \ref{appendix_rho}.

We solve the dipolar interaction following similar steps. First, note that, despite the decaying factor $r_{ij}^{-3}$,
the isotropic contribution is identical to the exchange term, while the anisotropic dipolar contribution is written as
\begin{IEEEeqnarray}{l}
\label{eq.dip_anisotropic}
({\bf S}_i\cdot{\bf r}_{ij})({\bf S}_j\cdot{\bf r}_{ij})=2z_{ij}y_{ij}\sqrt{S^2-(S_i^z)^2}S_j^z\sin\varphi_i+\nonumber\\
+\sqrt{S^2-(S_i^z)^2}\sqrt{S^2-(S_j^z)^2}(x_{ij}^2\cos\varphi_i\cos\varphi_j+\nonumber\\
+y_{ij}^2\sin\varphi_i\sin\varphi_j)+z_{ij} S_i^zS_j^z
\end{IEEEeqnarray}
where the mixed terms $S_i^x S_j^y$ and $S_i^x S_j^z$ were omitted since they do not yield second-order contributions.
We then expand the trigonometric functions as $\cos\varphi\approx 1-\rho_d\varphi^2/2$ and $\sin\varphi\approx\sqrt{\rho_d}\varphi$,
where $\rho_d$ is the dipolar renormalization factor, given by
\begin{equation}
\label{eq.rho_d}
\rho_d=\frac{1}{2}\left[1-\frac{\langle (S^z)^2\rangle_0}{S^2}\right]\left[1+\exp(-2\langle\varphi^2\rangle_0)\right].
\end{equation}
Again, the average is determined using the harmonic Hamiltonian and details about the demonstration of the above equation
can be found in Appendix (\ref{appendix_rho}). Therefore, the quadratic dipolar Hamiltonian is written as
\begin{widetext}
\begin{IEEEeqnarray}{rCl}
H_\textrm{dip}^{(2)}&=&\frac{\mu_0}{4\pi}\frac{(g\mu_B)^2}{2}\sum_{ij}\left[\rho_d S^2\frac{1}{r_{ij}^3}\left(1-\frac{3y_{ij}^2}{r_{ij}^2}\right)\varphi_i\varphi_j-\rho_d S^2\frac{1}{r_{ij}^3}\left(1-\frac{3x_{ij}^2}{r_{ij}^2}\right)\varphi_i\varphi_i+\frac{1}{r_{ij}^3}\left(1-\frac{3z_{ij}^2}{r_{ij}^2}\right)S_i^z S_j^z-\right.\nonumber\\
&&\left.-\frac{1}{r_{ij}^3}\left(1-\frac{3x_{ij}^2}{r_{ij}^2}\right)S_i^z S_i^z-6S\sqrt{\rho_d}\frac{z_{ij}y_{ij}}{r_{ij}^2}\varphi_i S_j^z\right]
\end{IEEEeqnarray}
\end{widetext}
We must be careful with the $q=0$ limit in performing the Fourier transform, which presents a slow convergence. Only for 
finite values of $q$, the result is almost independent of the sample surface. We follow the standard procedures to 
properly evaluate the lattice sum in $q=0$ \cite{pr58.1098,pr99.1128}. First, we divide the sum into two regions: one inside a small sphere 
containing only the nearest sites and the other involving the entire sample outside the small sphere. For the cubic lattice,
the sum inside the small sphere vanishes, and the second sum can be converted into a volume integral, which is then
written as two surface integrals. The inner surface integral (over the small spherical surface) provides the
Lorentz factor $4\pi/3$, while the outer surface results in the demagnetization factor, which depends on the domain shape. 
Here, we consider an elongated ellipsoid whose major axis is along the magnetization direction, and thereby we can neglect 
the demagnetizing field. For $q>0$, the sum can be converted in an integral over the entire domain, which results in
the well-known Fourier transform \cite{pr99.1128}
\begin{IEEEeqnarray}{rCl}
D_{ab}(q)&=&\sum_r \frac{1}{r^3}\left(\delta_{ab}-\frac{3r_a r_b}{r^2}\right)e^{i{\bf q}\cdot{\bf r}}\nonumber\\
&=&\frac{4\pi}{v_\textrm{ws}}\left(\frac{q_a q_b}{q^2}-\frac{\delta_{ab}}{3}\right)+\mathcal{O}(q^2),
\end{IEEEeqnarray}
where $v_\textrm{ws}=a^3$ is the volume of a cubic Wigner-Seitz cell with edge $a$ (lattice parameter). 
In momentum space, the quadratic dipolar Hamiltonian is then given by
\begin{IEEEeqnarray}{rCl}
\label{eq.hamiltoniandip}
H_\textrm{dip}^{(2)}&=&\mu_0\frac{g\mu_B M}{2S}\sum_q\left[\rho_d S^2 \frac{q_y^2}{q^2}\varphi_q\varphi_{-q}+\frac{q_z^2}{q^2}S_q^z S_{-q}^z+\right.\nonumber\\
&&\left.+2\sqrt{\rho_d}\frac{q_y q_z}{q^2}\varphi_q S_{-q}^z\right],
\end{IEEEeqnarray}
where $M=g\mu_B S/v_\textrm{ws}$ is the magnetization. Note that there is no dependence on $q_x$ and the dipolar 
contribution vanishes for vector momentum along the $\langle 100\rangle$ direction (considered as the preferred magnetization direction).
This implies that along the magnetization direction, spin-waves have lower energy since there is no effective dipolar field influence in this
direction.  

\subsection{Semiclassical approach}
Before we quantize the Hamiltonian, let us analyze the semiclassical limit.
Gathering both exchange and dipolar contributions, we obtain the harmonic Hamiltonian
\begin{equation}
\label{eq.H0}
H_0=\sum_q\left[S^2 A_q \varphi_q\varphi_{-q}+B_q S_q^z S_{-q}^z+ SC_q\varphi_q S_{-q}^z\right],
\end{equation}
where we have defined the coefficients
\begin{IEEEeqnarray}{rCl}
\IEEEyesnumber
\IEEEyessubnumber*
A_q&=&zJ\rho(1-\gamma_q)+J_d\rho_d\sin^2\theta_q\cos^2\phi_q\\
B_q&=&zJ(1-\gamma_q)+J_d\sin^2\theta_q\sin^2\phi_q\\
C_q&=&2J_d\sqrt{\rho_d}\sin^2\theta_q\sin\phi_q\cos\phi_q,
\end{IEEEeqnarray}
with $J_d=(\mu_0 g\mu_B M)/2S$. In addition, $\theta_q$ and $\phi_q$ are the polar and azimuth angles between ${\bf q}$
and the preferred magnetization direction, defined by $q_y=q\sin\theta_q\cos\phi_q$ and $q_z=q\sin\theta_q\sin\phi_q$. 
Note that momentum components appear only in a quadratic form in the Hamiltonians (\ref{eq.hamiltonianexc}) and (\ref{eq.hamiltoniandip}). 
Hence, the $A_q$, $B_q$, and $C_q$ coefficients are symmetric under the replacement ${\bf q}\to-{\bf q}$.
As expected, the dipolar energy vanishes along the magnetization direction and reaches its maximum for 
$\theta_q=\pi/2$. The spin-wave dynamics is obtained from the Hamilton equations
\begin{IEEEeqnarray}{rCl}
\hbar\dot{S}_q^z&=&-\frac{\partial H_0}{\partial\varphi_q}=-2SA_q S_q^y-SC_q S_q^z\\
\hbar\dot{S}_q^y&=&S\frac{\partial H_0}{\partial S_q^z}=2SB_q S_q^z+SC_q S_q^y,
\end{IEEEeqnarray}
where we use $S_q^y\approx S\varphi_q$. Thus, considering solutions in the form of harmonic travelling spin-waves 
with temporal dependence proportional to $\exp(i\omega_q t)$, we obtain the system of linear equations 
involving the transverse components
\begin{equation}
\left(\begin{array}{cc}
i\hbar\omega_q-SC_q & -2SB_q \\
2SA_q & i\hbar\omega+SC_q \\
\end{array}\right)\left(\begin{array}{c}
S_q^y\\
S_q^z\\
\end{array}\right)=0,
\end{equation}
which provides the spin-wave energy 
\begin{equation}
\label{eq.omega}
E_q=\hbar\omega_q=S\sqrt{4A_qB_q-C_q^2}.
\end{equation}

For the pure exchange Hamiltonian, it is easy to verify that $E_q=2SJz\sqrt{\rho}(1-\gamma_q)$, and so
we can define the renormalized exchange coupling as $J_r(T)=\sqrt{\rho(T)}J$. For the dipolar model,
it is not possible to exactly factorize the expression and obtain the same result; however, since $\rho\approx\rho_d$,
we can still use $J_r(T)$ as an approximated result.

Using the harmonic Hamiltonian given by Eq. (\ref{eq.H0}), we can evaluate the required statistical averages for
determining the renormalization factors, to wit
\begin{equation}
\langle S_q^zS_{-q}^z\rangle_0=\frac{2S^2A_q}{\beta E_q^2}
\end{equation}
and
\begin{equation}
\langle\varphi_q\varphi_{-q}\rangle_0=\frac{2B_q}{\beta E_q^2},
\end{equation}
with $\beta=(k_B T)^{-1}$. In above equations, we extend the limit of integration to $\pm\infty$ due to the fast 
decreasing of $\exp(-\beta H_0)$ at low temperatures. The distributions in the partition function are then treated as Gaussian.
\begin{figure}[h]
\centering \epsfig{file=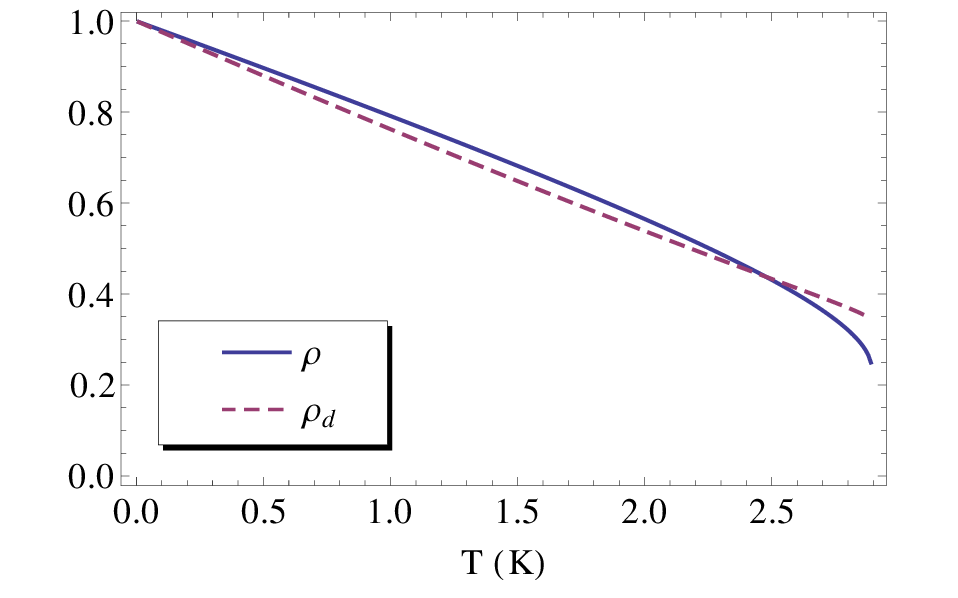,width=0.9\linewidth}
\caption{The renormalization factors for $S=1$, $J/k_B=1$ K, and $J_d=0.1 J$. At $T_c=2.90$K, $\rho$ abruptly drops to zero while
$\rho_d$ assumes a finite value.}
\label{fig.rho_cl}
\end{figure}

The renormalization parameters are determined by solving the $\rho$ and $\rho_d$ equations self-consistently. 
In general, the convergence is very fast and are required few iterations to reach the solution.
At the zero-temperature limit, it is easy to verify that $\rho$ and $\rho_d$ tend to one, as one sees in Fig. (\ref{fig.rho_cl}),
which allows to simplify the energy equation and obtain the spin-wave dispersion relation 
\begin{equation}
\label{eq.dispersion_cl}
\omega_q=\frac{2JSz}{\hbar}(1-\gamma_q)\left[1+\frac{\hbar\omega_M\sin^2\theta_q}{2JSz(1-\gamma_q)}\right]^{1/2},
\end{equation}
where $\omega_M=g\mu_B\mu_0 M$. The above outcome coincides with the well-known result for a ferromagnetic 
insulator with dipolar interaction \cite{rezende}. Indeed, one can determine the dispersion relation by using
the Landau-Lifschitz equation $\dot{\bf M}=-\gamma\mu_0{\bf M}\times{\bf H}_\textrm{eff}$, where the effective field
${\bf H}_\textrm{eff}={\bf H}_\textrm{exc}+{\bf H}_\textrm{dip}$. In the Fourier space, the exchange and dipolar fields 
[considering the magnetostatic limit for which $\nabla\cdot({\bf M}+{\bf H}_\textrm{dip})=0$, and $\nabla\times{\bf H}_\textrm{dip}={\bf 0}$]
are respectively given by ${\bf H}_\textrm{exc}=z\kappa {\bf M}_q-\kappa (aq)^2{\bf M}_q^\perp$, and 
${\bf H}_\textrm{dip}=-(\hat{{\bf q}}\cdot{\bf M}_q^\perp)\hat{{\bf q}}$, where the transverse
magnetization is ${\bf M}_q^\perp={\bf M}_q^y+{\bf M}_q^z$, and $\kappa=2JS/\hbar\omega_M$. Note that we have considered a 
uniform longitudinal magnetization component and the vector ${\bf q}$ was defined as previously. 
Adopting the oscillating time behavior for the magnetization, ${\bf M}_q(t)={\bf M}_q(0)e^{i\omega_q t}$, the Landau-Lifschitz 
equation yields
\begin{equation}
\omega_q=\frac{2JS q^2}{\hbar}\left[1+\frac{\hbar\omega_M\sin^2\theta_q}{2JS q^2}\right]^{1/2},
\end{equation}
which is the long wavelength limit of Eq. (\ref{eq.dispersion_cl}).
Considering $S=1$, $J/k_B=1$ K, and $J_d/k_B=0.1$ K, we determine the critical temperature (the point in which $\rho$ abruptly vanishes) 
at $T_c=2.90$ K. Opposite to the exchange renormalization factor, $\rho_d$ is finite at $T_c$. For $T>0$, the renormalization factors 
are slightly different; however, for $T\lesssim 0.9T_c$, $\rho\approx\rho_d$ and the dispersion relation can be written as 
$\omega_q(T)\approx \sqrt{\rho(T)}\omega_q(T=0)$.

The magnetization along the $\langle 100\rangle$ direction is readily determined and given by
\begin{equation}
\langle S^x\rangle = S-\frac{1}{N}\sum_q\left(\frac{1}{2S}\langle S_q^zS_{-q}^z\rangle_0+\frac{S}{2}\langle\varphi_q\varphi_{-q}\rangle_0\right),
\end{equation}
while $\langle S^z\rangle$ and $\langle S^y\rangle$ are null. 

\subsection{Quantum Hamiltonian}
Although the semiclassical results are in reasonable agreement with the literature, we can get better outcomes through 
the quantized Hamiltonian. To perform the quantization of the spin-waves, we promote the fields $\varphi_q$ and $S_q^z$ to
operators that satisfy $[\varphi_q,S_{q^\prime}^z]=i\delta_{qq^\prime}$. Therefore, it is convenient to
define the bosonic operators $a_q$ expressed by
\begin{IEEEeqnarray}{rCl}
\label{eq.phiSz_a}
\IEEEyesnumber
\IEEEyessubnumber*
\varphi_q&=&\frac{1}{\sqrt{2S}}(a_q^\dagger+a_{-q})\\
S_q^z&=&\frac{i\sqrt{S}}{\sqrt{2}}(a_q^\dagger-a_{-q}),
\end{IEEEeqnarray}
which leads to
\begin{IEEEeqnarray}{rCl}
H_0&=&\sum_q\frac{S}{2}\left[(A_q+B_q)(a_q^\dagger a_q+a_{-q}a_{-q}^\dagger)+(A_q-B_q+\right.\nonumber\\
&&\left.+iC_q)a_q^\dagger a_{-q}^\dagger+(A_q+B_q-iC_q)a_{-q}a_q\right].
\end{IEEEeqnarray}
The diagonalization is obtained from the definition of new bosonic operators by the Bogoliubov
transform
\begin{equation}
\label{eq.bogoliubov}
a_q=e^{i\psi_q/2}\cosh\chi_q \alpha_{-q}+e^{i\psi_q/2}\sinh\chi_q \alpha_q^\dagger,
\end{equation}
where $\chi_q$ is established by the relation
\begin{equation}
\tanh\chi_q=-\frac{|A_q-B_q+iC_q|}{A_q+B_q}
\end{equation}
and $\psi_q$ is the phase of $(A_q-B_q+iC_q)$. Therefore, after a straightforward procedure, we obtain
\begin{equation}
\label{eq.H0_quantum}
H_0=\sum_q \hbar\omega_q \left(\alpha_q^\dagger \alpha_q+\frac{1}{2}\right),
\end{equation}
where the spin-wave energy is again given by the relation $\hbar\omega_q=\tilde{S}\sqrt{4A_qB_q-C_q^2}$, and
we replace the classical spin value $S$ by $\tilde{S}=\sqrt{S(S+1)}$.

The renormalization parameters are calculated using the same Eqs. (\ref{eq.rho}) and (\ref{eq.rho_d}) but
the averages are determined by using the quantum harmonic Hamiltonian. Through the relation between spin and
bosonic operators, as well as the Bogoliubov transform, we are able to achieve
\begin{equation}
\langle\varphi_q\varphi_{-q}\rangle_0=\frac{B_q}{E_q}\coth\left(\frac{\beta E_q}{2}\right)
\end{equation}
and
\begin{equation}
\langle S_q^z S_{-q}^z\rangle_0=\frac{S^2A_q}{E_q}\coth\left(\frac{\beta E_q}{2}\right).
\end{equation}
Note that, in the small energy limit, {\it i.e.}, $E_q\ll k_B T$, we approximate $\coth(\beta E_q/2)$ by
$2/\beta E_q$, and the semiclassical averages are recovered.

\begin{figure}[h]
\centering \epsfig{file=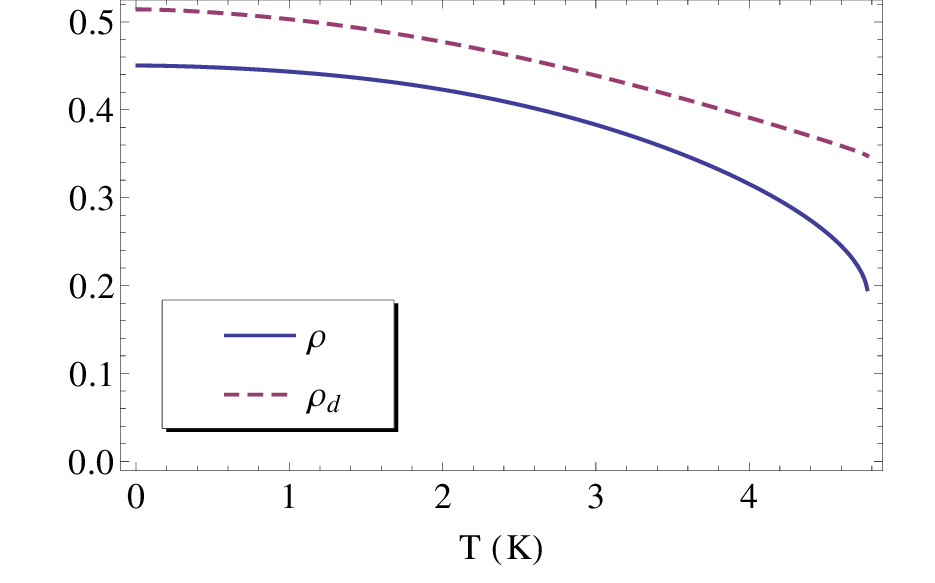,width=0.9\linewidth}
\caption{The renormalization factors for $\tilde{S}=\sqrt{2}$, $J/k_B=1$ K, and $J_d=0.1 J$. Here, the critical temperature is
given by $T_c=4.78$ K.}
\label{fig.rho_quantum}
\end{figure}

Fig. (\ref{fig.rho_quantum}) shows the temperature dependence of the renormalization factors. Due to the replacement of
$S$ by $\tilde{S}$, we observe an increasing in the critical temperature, from $T_c=2.90$ K to $T_c=4.78$ K. As we will
see in the next section, the quantum results better agree with experimental measures.
In addition, as one can see, $\rho_d\gtrsim \rho$, and both parameters are smaller than the respective 
semiclassical results. As expected, magnons have a more expressive contribution to disorder the 
magnetic phase in the quantum limit.

\begin{figure}[h]
\centering \epsfig{file=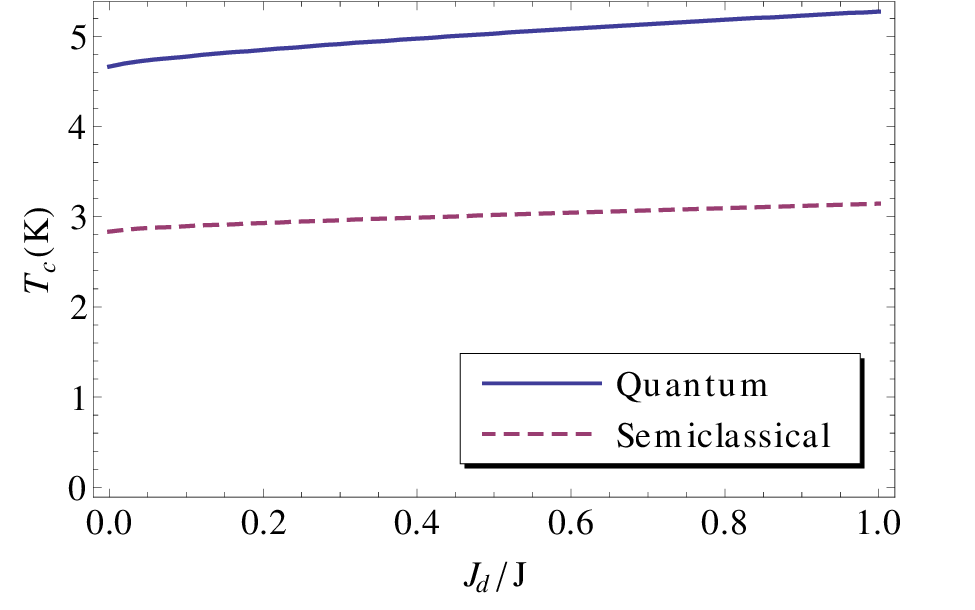,width=0.9\linewidth}
\caption{The critical temperature as function of the dipolar interaction intensity.}
\label{fig.Tc_Jd}
\end{figure}

The effect of the dipolar interaction on the critical temperature is shown in Fig. (\ref{fig.Tc_Jd}. Using the
semiclassical approach, we obtain $T_c$ almost constant for $0\leq J_d\leq J$, while for the quantum analysis, 
$T_c$ is a slightly increasing function of $J_d$. Despite the small correction, the dipolar interaction is important
to determine the transition temperature in Europium Chalcogenides with more precision, as demonstrated in the next section. 

\section{Europium Chalcogenides}
To verify the effectiveness of the SCHA including dipolar interactions, we apply the formalism
to determine the energy spectrum and transition temperature of Europium Oxide (EuO) and Europium Sulfite (EuS), two 
ferromagnetic insulators with well-known properties \cite{ibm14.214,prb12.2844,prb14.4897,prb14.4908,prb14.4923,wachter,prb22.5447,prb30.6504,prb47.11962,hasegawa}. 

The Europium Chalcogenides present a bulk with a high degree of symmetry that makes them ideal to be described by
the Heisenberg Hamiltonian. The magnetic properties arise from Eu$^{2+}$ ions that exhibit spin $S=7/2$ and orbital angular
momentum $L=0$. Because of the highly localized wave-function of the 4f electrons, the exchange coupling between Eu$^{2+}$
ions occurs due to indirect exchange with unoccupied 5d conduction bands and superexchange coupling \cite{ibm14.214}. 
The Europium ions are localized in an fcc lattice with isotropic exchange interaction between nearest and
near-nearest neighbors as well as long-range dipolar interaction. Neither the exchange nor the dipolar interaction shows
a preferred magnetization direction; however, a small anisotropy from the crystalline field orients the spin on the direction
$\langle 111\rangle$. Note that the Heisenberg Hamiltonian and the isotropic part of the dipolar interaction are invariant under
rotations, and they show the same structure whether we orient the magnetization in $\langle 100\rangle$ or 
$\langle 111\rangle$ directions. On the other hand, we must take care of the anisotropic part of the dipolar interaction 
when we rotate the preferred magnetization axis. Here, we work with directions according to the fcc lattice, and 
the anisotropic part of the dipolar interaction is adjusted to this frame (initially it is written in such a way that the x-axis
coincides with the $\langle 111\rangle$ direction and later it is rotated to a new frame where the x-axis coincides with 
the $\langle 100\rangle$ direction). The Hamiltonian is written as
\begin{IEEEeqnarray}{l}
\label{eq.hamiltonian2}
H=-J_1\sum_{\langle ij\rangle}{\bf S}_i\cdot{\bf S}_j-J_2\sum_{\langle\langle ij\rangle\rangle}{\bf S}_i\cdot{\bf S}_j+\nonumber\\
+\frac{\mu_0}{4\pi}\frac{(g\mu_B)^2}{2}\sum_{ij}\left[\frac{{\bf S}_i\cdot{\bf S}_j}{r_{ij}^3}-3 \frac{({\bf S}_i\cdot{\bf r}_{ij})({\bf S}_j\cdot{\bf r}_{ij})}{r_{ij}^5}\right],
\end{IEEEeqnarray}
where $J_1$ and $J_2$ represent the exchange coupling between nearest (nn) and near-nearest (nnn) neighbors,
respectively. For EuO ($a=5.141 \si{\angstrom}$), we adopt $J_1/k_B=0.606$ K and $J_2/k_B=0.119$ K, while $J_1/k_B=0.236$ K and 
$J_2/k_B=-0.118$ K are the exchange couplings for EuS ($a=5.960 \si{\angstrom}$) \cite{prb14.4897}. Using the SCHA, we obtain a harmonic 
Hamiltonian similar to the Eq. (\ref{eq.H0}); however, the coefficients are now given by
\begin{IEEEeqnarray}{rCl}
\IEEEyesnumber
\IEEEyessubnumber*
A_q&=&z_1J_1\rho_1(1-\gamma_q^{(nn)})+z_2J_2\rho_2(1-\gamma_q^{(nnn)})+\nonumber\\
&&+J_d\rho_d\sin^2\theta_q\cos^2\phi_q\\
B_q&=&z_1J_1(1-\gamma_q^{(nn)})+z_2J_2\rho(1-\gamma_q^{nnn})+\nonumber\\
&&+J_d\sin^2\theta_q\sin^2\phi_q\\
C_q&=&2J_d\sqrt{\rho_d}\sin^2\theta_q\sin\phi_q\cos\phi_q,
\end{IEEEeqnarray}
where the coordination numbers are $z_1=12$ and $z_2=6$, and the structure factors are
\begin{IEEEeqnarray}{rCl}
\IEEEyesnumber
\IEEEyessubnumber*
\gamma_1^{(nn)}&=&\frac{1}{6}\left[\cos\left(\frac{q_x}{2}\right)\cos\left(\frac{q_y}{2}\right)+\cos\left(\frac{q_y}{2}\right)\cos\left(\frac{q_z}{2}\right)\right.+\nonumber\\
&&+\left.\cos\left(\frac{q_z}{2}\right)\cos\left(\frac{q_x}{2}\right)\right],\\
\gamma_q^{(nnn)}&=&\frac{1}{3}(\cos q_x+\cos q_y +\cos q_z).
\end{IEEEeqnarray}
The angles $\theta_q$ and $\phi_q$ are defined in relation to the magnetization preferred axis ($\langle 111\rangle$ direction) 
and, in the reference frame used to define $\gamma_q^{(nn)}$ and $\gamma_q^{(nnn)}$, we have
\begin{equation}
\cos\theta_q=\frac{q_x+q_y+q_z}{\sqrt{3}q},\ \ \tan\phi_q=\frac{q_x+q_y-2q_z}{\sqrt{3}(q_x-q_y)}.
\end{equation}

\begin{figure}[h]
\centering \epsfig{file=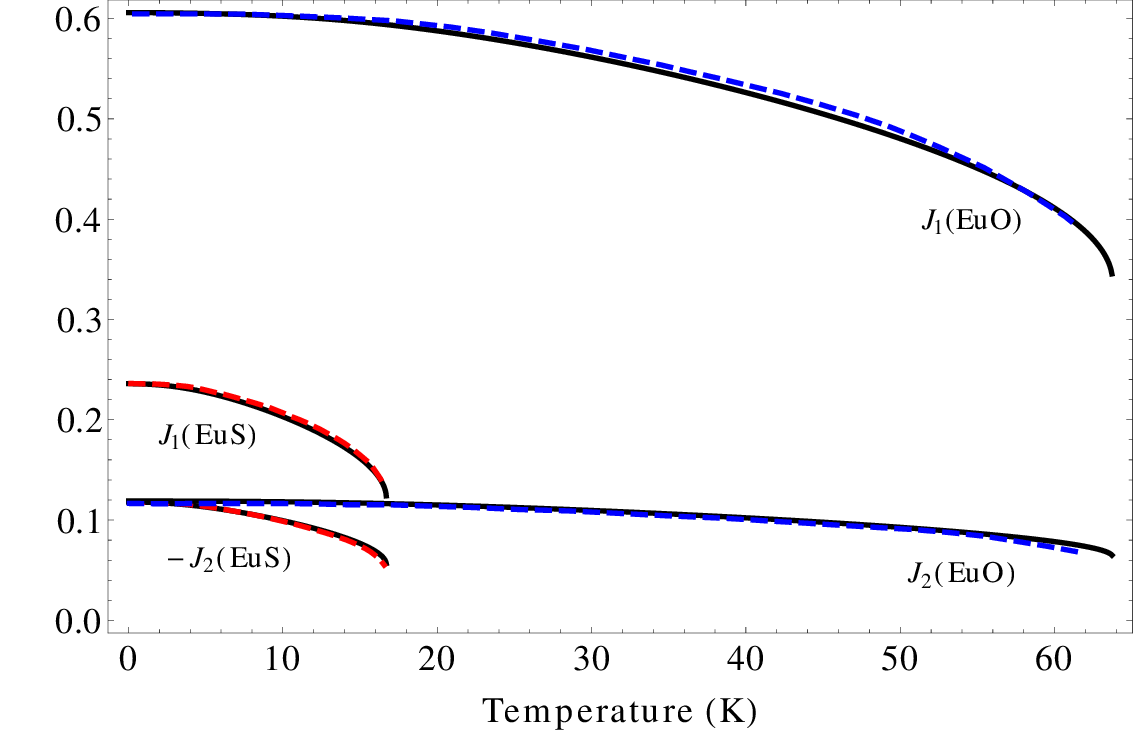,width=0.9\linewidth}
\caption{The renormalized coupling $J_1$ and $J_2$ (in units of $k_B$) as function of temperature for 
EuO and EuS. The black solid lines are the SCHA results, the dashed red line is from Ref. \cite{prb14.4923}, 
and the dashed blue line is from Ref. \cite{prb30.6504}}.
\label{fig.J1J2_T}
\end{figure}

Here, we consider three renormalization factors, again evaluated using Eqs. (\ref{eq.rho}) and (\ref{eq.rho_d}),
one for each contribution of the Hamiltonian. For the exchange coupling, the structure factor $\gamma_q$ is
chosen according to the nn or nnn interaction. As in the pure exchange Hamiltonian, we define the renormalized coupling terms 
by $J_1(T)=\sqrt{\rho_1(T)}J_1(0)$ and $J_2(T)=\sqrt{\rho_2(T)}J_2(0)$, which reproduce very well the result 
obtained from Dyson-Maleev representation for EuO \cite{prb14.4923} and EuS \cite{prb30.6504}. Both quantum SCHA and Dyson-Maleev
results are shown in Fig. (\ref{fig.J1J2_T}). The solid line ends at the critical temperature (the point where the
renormalization parameter abruptly vanishes), very close to the Curie temperature. Applying the quantum SCHA,
we obtain $T_c=63.75 K$ for EuO and $T_c=16.71$ K for EuS, while the experimental values of the Curie temperatures are
$T_C=69.15$ K (EuO) and $T_C=16.57$ K (EuS) \cite{prb14.4897}. The transition temperatures evaluated from both semiclassical
and quantum SCHA for the pure exchange Hamiltonian and the full model, including dipolar interaction, are listed in 
Table (\ref{table.tc}). As one can see, we reach the best results by using the quantum SCHA including the dipolar interaction.
\begin{figure}[h]
\centering \epsfig{file=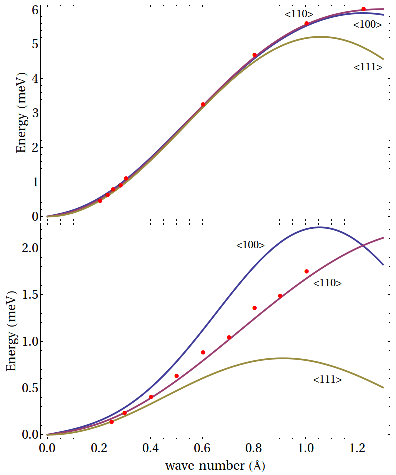,width=0.9\linewidth}
\caption{Spin-wave dispersion relation (at $T=5.5$ K) for EuO (above) and EuS (below). 
The solid lines are results of the SCHA formalism, and the red points are neutron scattering measures 
along the $\langle 110\rangle$ direction extracted from Ref. \cite{prb14.4897}.}
\label{fig.dispersion}
\end{figure}

Using the SCHA formalism, we also evaluate the spin-wave dispersion at $T=5.5$ K to compare with neutron scattering
experiments \cite{prb14.4897}. Fig. (\ref{fig.dispersion}) shows the curves (solid lines) calculated through SCHA and the 
experimental data for EuO and EuS. In both cases, the SCHA provides good agreement with the experiments. In addition,
Fig. ({\ref{fig.magnetization}) shows the temperature dependence of the reduced magnetization for both EuO (in blue) and EuS (in red). 
The figure plots the data from neutron scattering experiments \cite{prb14.4908} (following the same color 
scheme). For EuS (EuO), the calculated magnetization is slightly smaller (larger) than the experimental measurements. The
experimental measurements were done using thin-slab samples obtained from polycrystalline powders \cite{prb14.4897}. Surface 
effects, which were disregarded in our analysis, could justify the slight difference between the theoretical and experimental results.
For the same reason, one could justify the difference between the SCHA and experimental critical temperature, mainly for EuO.
Since exchange couplings are intrinsic properties, depending on the superposition of electron wave-functions, surface contributions 
are negligible for the renormalized coupling $J_1$ and $J_2$, which explain the good agreement in Fig. (\ref{fig.J1J2_T}). 
On the other hand, D. A. Garanin applied the SCGA with a simplified dipolar interaction to obtain a critical temperature
$T_c\approx 69 K$ for the EuO \cite{prb51.16413}. At the same time, the magnetization obtained by Garanin is slightly larger than the 
experimental data, as the SCHA result. Therefore, more investigation is necessary to explain the slightly lower EuO critical 
temperature obtained from SCHA.
 
\begin{table}
\centering
\begin{tabular}{l  l  l}
\hline\hline
Method 									& EuO  	  & EuS \\
\hline
Pure exchange, semiclassical SCHA 		& 48.64 K & 17.64 K \\
Pure exchange, quantum SCHA	 			& 61.24 K & 15.22 K \\
Full model, semiclassical SCHA 			& 50.60 K & 19.49 K \\
Full model, quantum SCHA					& 63.75 K & 16.71 K \\
Experimental data \cite{prb14.4897}	& 69.15 K & 16.57 K \\
\hline
\end{tabular}
\label{table.tc}
\caption{Transition temperature obtained from SCHA considering pure exchange Hamiltonian and including dipolar
interaction. Both semiclassical and quantum are considered and the best result is obtained by using the quantum
SCHA with dipolar interaction.}
\end{table}

\begin{figure}[h]
\centering \epsfig{file=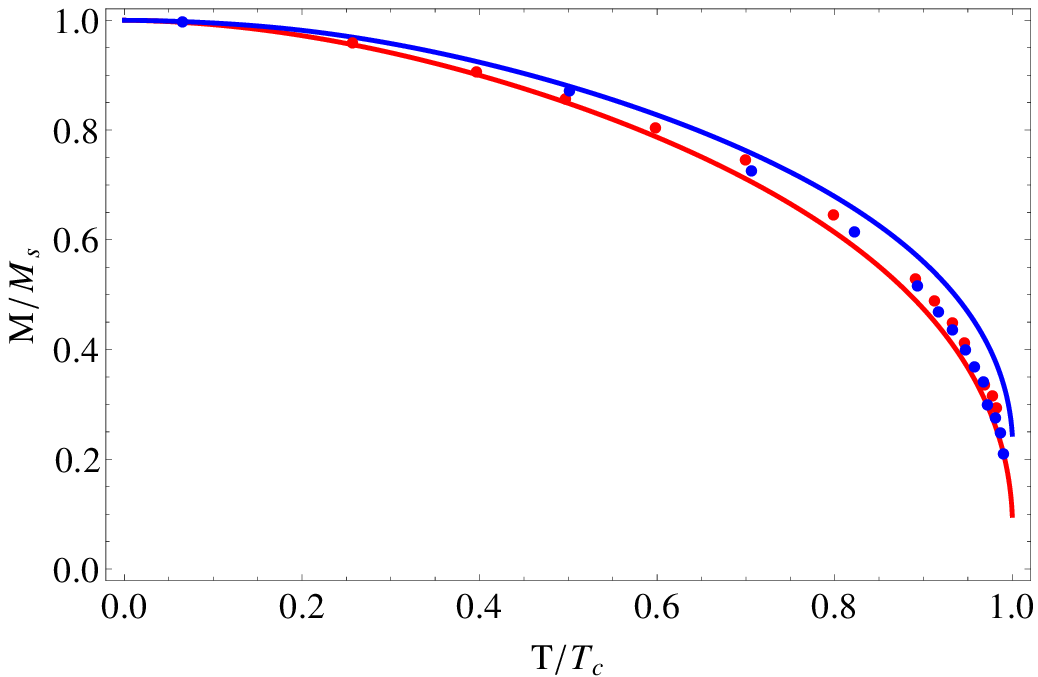,width=0.9\linewidth}
\caption{The reduced magnetization curve as function of relative temperature $T/T_c$ (EuO in blue and EuS in red). 
The solid lines are the SCHA results and the points were obtained for neutron scattering experiments \cite{prb14.4908}}
\label{fig.magnetization}
\end{figure}

\section{Summary and conclusions}
The SCHA method is an efficient formalism for treating spin models at finite temperature, provided that the method deals 
with simple harmonic Hamiltonians. The temperature dependence is included through renormalization parameters that need to be 
solved by self-consistent equations. The SCHA has been used for many years, mainly to determine critical 
temperatures; however, in general, only short-range interaction has been considered. Since dipolar interaction is
a relevant contribution for many magnetic experiments, we developed the SCHA to include the dipolar field. 

Using the SCHA, we obtained the spin-wave spectrum energy, the transition temperature, renormalized couplings, 
and magnetization, while other results can also be easily determined. The SCHA results coincide with theoretical development obtained 
from standard bosonic representations in the zero-temperature limit. To verify our equations, we 
applied the formalism to the ferromagnetic Europium Chalcogenides: EuO and EuS. The results were compared with well-known 
results obtained from usual bosonic representations and neutron scattering experiments. We achieved a good agreement for both 
materials, mainly for Europium Sulfite. For EuS, the transition temperature provided by SCHA (16.71 K) showed an error of less 
than 1 percent when compared to the experimental value (16.57 K). Besides, the theoretical magnetization curve presented an 
excellent fit with the neutron scattering data. In the EuO case, the determined transition temperature was $T_c=63.75$ K, while the
experimental result is given by $T_c=69.15$ K. It is known that anisotropic energy provides a small gap of the order of
3.5 $\mu$eV in EuO (and one order of magnitude smaller in EuS) that was disregarding in our work. However, 
this anisotropy is insufficient to explain the observed difference. In addition, surface effects, which were also disregarded, 
could contribute to the small difference between theoretical and experimental results. The calculated magnetization curve is 
precise at the low-temperature limit, but SCHA gives results slightly different from the experimental data for
$T\gtrsim 0.5 T_c$ (approximately 10$\%$ larger at $T=0.9 T_c$, for EuO). Finally, the renormalized exchange
coupling evaluated by SCHA for both materials showed an excellent agreement with the literature.

The SCHA revealed a great formalism in describing the magnetic thermodynamics in models endowed with dipolar 
interaction. Likewise the usual bosonic formalisms, the SCHA method can be applied in the theoretical investigation of 
magnetism in various fields. Since the SCHA formalism is a quadratic model, its application is more 
manageable than traditional bosonic representations that need to include spin-wave interaction to consider corrections 
of fourth-order or higher. In addition, the SCHA shows some advantages compared with the SCGA, which is another method
with similar purposes. The first one is related to the smaller number of required self-consistent parameters. For example, 
for the isotropic ferromagnetic Heisenberg model in the presence of a field $H^z$, the SCHA requires only a parameter 
(the renormalization term $\rho$) to determine the thermodynamics. 
At the same time, SCGA needs three self-consistent unknown variables (the magnetization and the 
longitudinal and transverse components of the mean-field fluctuation) to treat the same Hamiltonian. 
The difference in self-consistent parameters between both methods is even more significant for more general Hamiltonians. 
Another advantage refers to the quantization procedure. Whereas the SCGA applies to only classical spin models, the 
SCHA Hamiltonian, which is written in terms of $S^z$ and $\varphi$, can be easily quantized by imposing the commutation relation 
$[\varphi_i,S_j^z]=i\delta_{ij}$. Finally, despite the excellent
SCGA results, it is easier to implement the dipolar interaction in the SCHA than in the SCGA, which could be a decisive point
depending on the problem addressed.

\appendix
\section{Renormalization factors}
\label{appendix_rho}
The dynamics of $S^z$ depends on which Hamiltonian is used to determine the time derivative, and the result is strongly affected 
whether one uses the simple harmonic Hamiltonian (without any correction) or the complicated full version. Therefore, we include
the renormalization parameter in the quadratic Hamiltonian to improve the results without turning the evaluation overly laborious.

In order to determine the renormalization factors, we compare the average $\langle \dot{S}_q^z\dot{S}_{-q}^z\rangle_0$
evaluated using the quadratic Hamiltonian $H_0$, given by Eq. (\ref{eq.H0}), with the result obtained from the full 
Hamiltonian $H$, {\it i.e.} $\langle \dot{S}_q^z\dot{S}_{-q}^z\rangle$. Starting with the former, we obtain
\begin{IEEEeqnarray}{l}
\hbar^2\langle \dot{S}_q^z\dot{S}_{-q}^z\rangle_0=4S^4A_q^2\langle\varphi_q\varphi_{-q}\rangle_0+4S^3A_qC_q\langle\varphi_q S_{-q}^z\rangle_0+\nonumber\\
+S^2C_q^2\langle S_q^zS_{-q}^z\rangle_0=\frac{2S^2A_q}{\beta}.
\end{IEEEeqnarray}
To find out the second term, given by the Fourier transform
\begin{equation}
\langle \dot{S}_q^z\dot{S}_{-q}^z\rangle=\frac{1}{N}\sum_{ij}\langle \dot{S}_i^z\dot{S}_j^z\rangle e^{i{\bf q}\cdot({\bf r}_j-{\bf r}_i)},
\end{equation}
we use the following useful relation, obtained after an integration by parts,
\begin{IEEEeqnarray}{rCl}
\hbar^2\langle \dot{S}_i^z\dot{S}_j^z\rangle&=&\frac{1}{Z}\int\mathcal{D}\varphi\mathcal{D}S^z\frac{\partial H}{\partial\varphi_i}\frac{\partial H}{\partial\varphi_j}e^{-\beta H}\nonumber\\
&=&\frac{1}{Z}\int\mathcal{D}\varphi\mathcal{D}S^z\frac{1}{\beta}\frac{\partial^2 H}{\partial\varphi_i\partial\varphi_j}e^{-\beta H},
\end{IEEEeqnarray}
where $Z$ is the partition function, and the integration measure $\mathcal{D}\varphi\mathcal{D}S^z$ stands for the field integration 
over each site on the lattice. In addition, we extend the integration limit to $-\infty<\varphi,S^z<\infty$ and so we will deal 
with Gaussian integrals. Since the exchange and dipolar Hamiltonians are decoupled, we can develop them separately. 
For the exchange term, we get 
\begin{equation}
\frac{\partial^2 H_\textrm{exc}}{\partial\varphi_i\partial\varphi_j}=2\sum_\eta h_{i\eta}^\textrm{exc}(\delta_{j\eta}-\delta_{ij}),
\end{equation}
with $h_{ij}^\textrm{exc}=-J\sqrt{S^2-(S_i^z)^2}\sqrt{S^2-(S_j^z)^2}\cos(\varphi_i-\varphi_j)$ and $\eta$ being 
the nearest neighbors. Then, the Fourier transform yields
\begin{equation}
\hbar^2\langle \dot{S}_q^z\dot{S}_{-q}^z\rangle_\textrm{exc}=\frac{2zJ}{\beta}(1-\gamma_q)\langle h^\textrm{exc}\rangle_0,
\end{equation}
where we assume that $\langle h^\textrm{exc}\rangle_0$ is independent of the site position and the average is 
determined through the harmonic Hamiltonian $H_0$. We can evaluate the averages over $S^z$ and $\varphi$ independently
for decoupled fields. However, in our case, the Hamiltonian present mixed terms and, 
in space coordinates, it looks like
\begin{equation}
H_0=\sum_{ij}[S^2A_{ij}\varphi_i\varphi_j+B_{ij}S_i^z S_j^z+SC_{ij}\varphi_iS_j^z].
\end{equation}
Then, some extra steps are necessary to properly determine $\langle h^\textrm{exc}\rangle_0$. Firstly, we consider that 
$S_i^z\approx S_j^z\ll S$ and write $\langle h^\textrm{exc}\rangle_0=S^2\langle(1-s^z)\cos\phi\rangle$, where $s^z=S^z/S$ and 
$\phi=\Delta\varphi$. Since we are dealing with Gaussian integrals, we can replace $\langle\cos\phi\rangle_0$ by 
$\exp(-\langle\phi^2\rangle_0/2)$. Now, to determine the average $\langle (s^z)^2\cos\phi\rangle_0$, we expand the cosine 
function and apply the general result for Gaussian distributions, for example, in the variables $x$ and $y$
\begin{equation}
\langle x^{2m}y^{2n}\rangle=\sum_{p=0}^{\textrm{min}(m,n)}c_{mnp}\langle x^2\rangle^{m-p}\langle y^2\rangle^{n-p}\langle xy\rangle^{2p},
\end{equation}
whose coefficients are given by
\begin{equation}
c_{mnp}=\frac{(2m)!(2n)!(2m-2p-1)!!(2n-2p-1)!!}{(2m-2p)!(2n-2p)!(2p)!}
\end{equation}
Therefore, it is a straightforward procedure to get the exact result
\begin{equation}
\langle (s^z)^2\cos\phi\rangle_0=\left[\langle(s^z)^2\rangle_0-\langle s^z\phi\rangle_0^2\right]\exp\left[-\frac{\langle\phi^2\rangle_0}{2}\right]
\end{equation}
and disregarding the forth-order term $\langle s^z\phi\rangle_0^2$, we obtain Eq. (\ref{eq.rho}). Note that, in the decoupled 
fields case, we have $\langle s^z\varphi\rangle_0=0$, which provides the result obtained from the relation $\langle (s^z)^2\cos\phi\rangle_0=\langle (s^z)^2\rangle_0\langle\cos\phi\rangle_0$. In addition, one can use the same approach to 
implement higher-order contributions to the SCHA renormalization parameter, although the corrections are minor and do not 
provide reasonable changes in final results. The more relevant contribution to determining the transition temperature comes from 
the exponential term.

The dipolar renormalization parameter follows the same development. Considering the isotropic part 
and the terms present in Eq. (\ref{eq.dip_anisotropic}), the derivative of the dipolar Hamiltonian reads
\begin{IEEEeqnarray}{l}
\frac{\partial^2 H_\textrm{dip}}{\partial\varphi_i\partial\varphi_j}=\frac{J_d v_\textrm{ws}}{4\pi}\sum_\eta\left[ 2\xi_{i\eta}^{xx}(\delta_{j\eta}\sin\varphi_i\sin\varphi_\eta-\right.\nonumber\\
-\delta_{ij}\cos\varphi_i\cos\varphi_\eta)+2\xi_{i\eta}^{yy}(\delta_{j\eta}\cos\varphi_i\cos\varphi_\eta-\nonumber\\
\left.-\delta_{ij}\sin\varphi_i\sin\varphi_\eta)-\xi_{i\eta}^{yz}\sin\varphi_i\delta_{ij}\right],
\end{IEEEeqnarray}
where $\eta$ now represents any site on the lattice, not only the nearest neighbors as previously, and the $\xi$ coefficients are
expressed by
\begin{IEEEeqnarray}{l}
\xi_{ij}^{xx}=\sqrt{S^2-(S_i^z)^2}\sqrt{S^2-(S_j^z)^2}\left(\frac{1}{r_{ij}^3}-\frac{3x_{ij}^2}{r_{ij}^5}\right),\nonumber\\
\xi_{ij}^{yy}=\sqrt{S^2-(S_i^z)^2}\sqrt{S^2-(S_j^z)^2}\left(\frac{1}{r_{ij}^3}-\frac{3y_{ij}^2}{r_{ij}^5}\right),\nonumber\\
\xi_{ij}^{yz}=-6\sqrt{S^2-(S_i^z)^2}S_j^z\frac{y_{ij}z_{ij}}{r_{ij}^5}.
\end{IEEEeqnarray}

For determining the averages of above equation, we make the same previous considerations. Besides, we adopt that the sine terms are 
much smaller than the cosine ones, provided that $\varphi\ll 1$. In performing the Fourier transform, we obtain
\begin{equation}
\hbar^2\langle \dot{S}_q^z\dot{S}_{-q}^z\rangle_\textrm{dip}=\frac{2J_d}{\beta}\langle(S^2-(S^z)^2)\cos^2\varphi\rangle_0\frac{q_y^2}{q^2},
\end{equation}
and comparing with $\hbar^2\langle\dot{S}_q^z\dot{S}_{-q}^z\rangle_0$, we see that the dipolar renormalization parameter 
is given by
\begin{equation}
\rho_d=\left\langle\left(1-\frac{(S^z)^2}{S^2}\right)\cos^2\varphi\right\rangle_0.
\end{equation}
After a fast algebraic manipulation, and using the same procedures of the exchange case, we finally 
obtain Eq. (\ref{eq.rho_d}). Other cases, as the near-nearest exchange interaction in the Europium Chalcogenides, are solved
following the same steps presented here.

\bibliography{manuscript}
\end{document}